\documentstyle[12pt,aaspp,flushrt]{article}

\begin{document}

\title{Gamma Ray Pulsars: Emission from Extended Polar Cap Cascades}

\author{Joseph K. Daugherty}
\affil{Department of Computer Science, University of North Carolina
at Asheville, Asheville, NC 28804}

\author{Alice K. Harding}
\affil{Laboratory for High Energy Astrophysics,
NASA/Goddard Space Flight Center,
Greenbelt, MD 20771}

\begin{abstract}
\end{abstract}
\newpage
\section{INTRODUCTION}
%
%
%
%
In recent years the Compton Gamma Ray Observatory (CGRO) and other
instruments have provided major new discoveries and detailed observations
of isolated $\gamma$-ray pulsars, including the Crab (Nolan et al. 1993),
Vela (Kanbach et al. 1994), Geminga (Halpern \& Holt 1992, Bertsch et
al. 1992, Mayer-Hasselwander et al. 1994), PSR B1509-58 (Wilson et al. 1992),
PSR B1706-44 (Thompson et al. 1992), PSR B1055-52 (Fierro et al. 1993),
and most recently PSR B1951+32 (Ramanamurthy et al. 1995).
Models of these objects must now account for a variety of detailed features
in the emission, especially from the most intense sources (Crab, Vela,
Geminga). Current models have in fact already encountered problems
in explaining how these sources can show both remarkable similarities
and puzzling variations in their light curves and phase-resolved energy
spectra. These difficulties are even more severe if models of the
$\gamma$- ray emission must also be consistent with the radiation observed
at radio, optical, and X-ray wavelengths. As the observational statistics
for the weaker sources improve, these theoretical challenges may become
even more formidable.

At present two general types of $\gamma$- ray pulsar models are popular
in the literature. The Polar Cap (PC) model, first proposed by Sturrock (1971)
and later investigated by numerous authors (see for example Ruderman
and Sutherland 1975, Harding 1981, Daugherty and Harding 1982, Arons 1983)
assumes that the emission is produced by electrons accelerated to high
energies just above the surface of a magnetized rotating neutron star (NS),
in the vicinity of the magnetic poles. In contrast, the Outer Gap model
(Cheng, Ho, and Ruderman 1986a, 1986b) places the acceleration regions
much higher in the NS magnetosphere, in vacuum gaps formed within
a charge-separated plasma.

In a previous paper (Daugherty and Harding 1994, hereafter DH94)
we proposed a version of the PC model based on the following
principal assumptions:

(a) The gamma emission is initiated by the acceleration of electrons
from the NS surface, just above the magnetic PC regions
which enclose the {\it open\/} magnetic field lines extending to the
velocity-of-light cylinder (LC).

(b) The emission originates as curvature radiation (CR) produced by
the electrons as they follow the curvature of the open magnetic
field lines.

(c) The processes of direct $1-\gamma$ pair conversion (see for example
Erber 1966) by the NS magnetic field and synchrotron radiation (SR) by
the emitted pairs produce photon-pair cascades, from which the observed
$\gamma$ radiation emerges.

(d) The rotational and magnetic axes of the radiating NS are nearly
aligned, so that the inclination $\alpha$ is small enough to be comparable
with the PC half-angle $\theta_{pc}$. More precisely, the model requires
that $\alpha \sim \theta_b$ where $\theta_b$ is the half-angle of the
$\gamma$-beam emerging from the PC.

Assumptions (a)-(c) comprise essentially the original postulates of the
PC model (Sturrock 1971). They describe the overall physics of
the cascade process and in combination they determine the form of the
production spectra for the gamma rays and the pairs. The final assumption
(d) primarily affects the viewing geometry. It implies that randomly
oriented observers should see emission from at most one PC. However,
since CR-induced cascades are intrinsically hollow-cone sources which
produce their most intense
emission near the PC rim, observers viewing a single PC may detect light
curves with either single or double peaks (DH94).
Sterner and Dermer (1994) independently noted a similar effect in a
model of PC cascades initiated by Comptonization rather than CR.
The assumption that $\alpha \sim \theta_b$ allows the phase separation
between double peaks to become large enough to match the observed values
($\sim 0.4$ for the Crab and Vela, $\sim 0.5$ for Geminga).

In the present work we refine assumption (d) by requiring {\it only\/} that
$\alpha \sim \theta_b$, not that $\alpha$ itself be necessarily small.
Hence in place of the Nearly Aligned Rotator (NAR) model described in
DH94, we consider here a more general Single Polar Cap (SPC) model.
In addition, we introduce a further assumption which allows
$\theta_b$ (and $\alpha$) to have significantly larger values
than $\theta_{pc}$ itself:

(e) the acceleration of the electrons occurs over an extended distance
above the PC surface, so that they reach their peak energies at heights
of a few NS radii. Above these heights, the acceleration is
cut off by an overlying force-free plasma.

In DH94 we neglected the height of the acceleration
region and simply supplied the electrons with an injection energy at the NS
surface, then traced their CR energy losses as they escaped outward along
field lines for which ${\bf E \cdot B} \sim 0$. We have since noted that
the assumption
of an extended acceleration region provides a solution to a serious
difficulty with our previous model, namely the ``observability'' problem.
This refers to the fact that if conventional estimates of PC dimensions
are accurate, $\gamma$-beams emitted by energetic electrons just above the NS
surface would be so small that there would be a low probability
($\lesssim 10^{-2}$)
that they could be detected by randomly oriented observers.
In our previous work we noted that the usual estimate for the PC radius
$R_{pc}$ may in fact be too small, although moderate increases in
$R_{pc}$ cannot by themselves resolve the observability problem.
However, the outward flaring of the magnetic field lines implies that
the half-angle $\theta_{b}$ of the (hollow) cascade $\gamma$-beams
increases rapidly with height above the NS surface. Thus the effect
of extending the acceleration zone up to heights of a few NS radii,
especially if combined with moderately increased ($\lesssim 2$)
PC dimensions, can produce rotating beams whose edges sweep
over a much larger solid angle.

In DH94 we also noted that in order to produce double
peaks as narrow as those observed from the Crab, Vela, and Geminga,
we had to assume that the surface density of the electrons drawn from
the NS surface is concentrated near the PC rim. In the present work
we suggest a physical basis for this empirical observation, namely
the acceleration of secondary cascade electrons created near the rim.
More precisely, the excess rim density may be supplied by a multistep
process initiated by the reversed acceleration of secondary positrons
created just below the acceleration cutoff height. These particles can
produce downward-oriented cascades, creating new pairs near the NS surface.
A fraction of the electrons from these pairs may then be accelerated
upward along with the true primary electrons, adding to the net outward
flow. We argue that this sort of cascade feedback process should occur
preferentially near the PC rim, where the open magnetic field lines have
their maximum curvature.

\newpage
\section{EXTENDED PAIR CASCADES}
%
%
%
%
%

In our treatment of the NAR Model in DH94, we assumed that
the acceleration of PC electrons starts at the NS surface and is cut off
sharply at a height $h \ll R_{pc}$ by an overlying force-free pair
plasma. This assumption was made partly for simplicity, and also
because there is still no firmly established, self-consistent
electrodynamical model for magnetospheric acceleration, either near the
NS surface or elsewhere. However, we note that a significant problem with
these models may be resolved if the accelerating potential $\Phi(h)$
extends upward to heights $h \sim 2 \-- 3$ NS radii or higher.
We will first discuss our motivation for exploring extended
acceleration regions, then describe our model results based on specific
empirical choices for $\Phi(h)$.

For simplicity we retain our NAR model assumption that each PC is almost
circular with radius $R_{pc} = R_{ns} \theta_{pc}$, where $R_{ns}$ is the
NS radius. While the more general SPC model allows larger values of
$\alpha$ and hence noncircular PCs, this approximation should be still
adequate for our present treatment. For purely dipolar fields,
the conventional estimate for the half-angle $\theta_{pc}$ is just
\begin{equation} \label{tpc}
\sin\theta_{pc} = \left({R_{ns} \over R_{lc}}\right)^{1/2}
= \left({R_{ns} \Omega\over c}\right)^{1/2}
\end{equation}
where $R_{lc} = c/\Omega$ denotes the distance to the velocity-of-light
cylinder and $\Omega$ is the NS angular rotation frequency. Eq. (\ref{tpc})
assumes that a dipole field line, emerging from a point near the PC rim,
should close just inside the light cylinder. However, as we noted in DH94
this estimate ignores all plasma effects and thus should be regarded only
as a lower limit on $\theta_{pc}$. For example, Michel (1982, 1991) has
found that the presence of a force-free, rigidly corotating plasma
(even without inertial effects or outward current flow) causes a
distortion of the field lines which increases the PC radius by a
factor $\sim 1.3$. Hence we argue that a more realistic model could be
expected to increase $\theta_{pc}$ by a factor $\sim 2$ over Eq. (\ref {tpc}).

If we make the usual assumption that the magnetic field is purely dipolar,
the equation describing a given field line emerging from the PC is just
\begin{equation} \label{dpl}
r = k \sin^{2}\theta
\end{equation}
where k is constant. At a given point on the field line, the angle $\psi$
of the local tangent (measured from the magnetic axis) is given by
\begin{equation} \label{tnl}
\tan\psi = {{3 \sin\theta \cos\theta} \over {3 {\cos^{2}\theta- 1}}}
\end{equation}
If the gamma beam size is approximately determined by the locus of tangents
to the outermost open field lines, for $\theta_{pc} \ll 1$ a cascade gamma
beam originating from the NS surface would have a half angle
$\theta_{b} \sim \tan\psi \sim {3 \over 2} \theta_{pc}$.

In general we can use Eqs. (\ref{dpl}) and (\ref{tnl}) to estimate
the increase in beam width $\theta_{b} \sim \psi$ with height, for a
given PC radius.
Figure 1 illustrates this height dependence by plotting the tangent angle
$\psi$ vs. radial distance along the field lines, for the case of the
Vela pulsar ($P = 0.89$ ms). The curves labeled 1, 2,... denote field
lines emerging from the NS surface at the corresponding multiples of
$\theta_{pc}$ as given by Eq. (\ref{tpc}). It is evident
that if the cascade gamma emission extends upward to heights exceeding
$\sim 3$ NS radii, $\theta_{b} \sim \psi$ can become significantly
larger than ${3 \over 2} \theta_{pc}$. This effect is even more
pronounced if $\theta_{pc}$ is taken to be $\sim 2$ or more times
the standard estimate (\ref{tpc}).

\section{ACCELERATION AND ENERGY LOSSES ABOVE POLAR CAPS}

We have shown that from the standpoint of viewing geometry, extended PC
cascades may provide a viable solution to the observability problem.
The obvious next step is to examine the possiblity that the acceleration
of electrons from the PC surface might be sustained up to heights of
several NS radii. This question also requires us to consider in detail
the energy loss mechanisms which may affect the net acceleration.

Due to the intense ($\sim 10^{12} G$) NS magnetic fields, electrons
accelerated from the PC surface are constrained by rapid SR losses
to follow the field lines. Hence they obey a one-dimensional equation
of motion, which may be expressed as an energy-balance equation:
\begin{equation} \label{erg}
{d\gamma \over ds} = (\beta c)^{-1}
\left[
  \left({d\gamma \over dt}\right)_{acc}
- \left({d\gamma \over dt}\right)_{cr}
- \left({d\gamma \over dt}\right)_{cs}
- \left({d\gamma \over dt}\right)_{other}
\right]
\end{equation}
Here $\gamma$ denotes the electron Lorentz factor, $\beta = v/c$, and
$s$ is the distance traversed along the field line. The subscripts
labeling the component energy gain and loss rates are defined as follows.

The subscript {\it acc\/} denotes the energy gain due to electrostatic
acceleration in regions where ${\bf E \cdot B}$ is nonzero.
We assume this term is proportional to ${\bf E_{\parallel}}$, the component
of ${\bf E}$ parallel to ${\bf B}$, at each point along the particle
trajectory (magnetic field line). Unfortunately, current models of
pulsar magnetospheres do not agree on the behavior of ${\bf E_{\parallel}(r)}$
near the PC surface. Hence the energy-gain term in Eq. (\ref{erg}) must
be regarded as unknown. However, we can at least assume various simple
models for the accelerating potential (e.g. Ruderman and Sutherland 1975,
Arons 1983) in our simulations and compare the results for each model
with observations. In Sections 6 and 7 we show that we have
been able to find self-consistent models of extended cascades which
yield light curves and spectra similar to the observed values. We have
also identified significant constraints on the accelerating field which
are critical to the viability of these models.

In contrast to the gain-rate term, the principal loss-rate terms in
Eq. (\ref{erg})are reasonably well
understood. The subscripts {\it cr\/}, {\it cs\/}, and {\it other\/}
denote energy losses due to CR, Compton upscattering (Dermer 1990,
Chang 1995), and other scattering processes respectively. One example
of the latter is triplet pair production (Mastichiadis et al. 1986,
Mastichiadis 1991, Dermer and Schlickeiser 1991). Sturner (1995) has
recently provided a systematic treatment of PC electron acceleration
which considers these energy-loss processes in detail. We have used
his expressions for the CS loss terms in our simulations, although his
treatment involves a number of simplifying approximations.

The CR loss rate has the simple form (see for example Jackson 1975)
\begin{equation} \label{crl}
\left({d\gamma \over dt}\right)_{cr} =
{2 \over 3} {e^2 \over {m c}}
{\gamma^4 \over {\rho_{c}}^2}
\end{equation}
where $\rho_{c}$ is the local radius of curvature of the magnetic field
line. For a purely dipolar field, the exact
expression for $\rho_{c}$ is just
\begin{equation} \label{crd}
\rho_{c} = {{k (\sin^4 \theta + \sin^2 2 \theta)}
\over
{\sin^4 \theta + 2 \sin^2 2 \theta - 2 \sin^2 \theta \cos 2 \theta}}
\end{equation}
Since Eq. (\ref{crd}) yields values $~\sim 10^7$ cm for standard
PC model parameters, the CR loss rate only becomes significant for
$\gamma \gtrsim 10^6$. At higher energies it is by far the dominant loss
mechanism.

The CS loss rate results from upscattering of ambient photons by the
accelerated electron beam. In our model the photon background consists of
thermal emission from the NS surface, and hence the CS loss rate should
only be significant only at heights $h \lesssim R_{ns}$ above the surface.
Pulsed X-ray observations of Geminga (Halpern and Ruderman 1993)
and PSR B1055-52 (\"Ogelman and Finley 1993)
suggest that for at least some sources the thermal
background may include multiple components at distinct temperatures
(e.g., emission from both the cooling NS surface and hotter
regions in the vicinity of the PCs).

The CS loss rate is found from the general expression (Dermer 1990,
Sturner 1995)
\begin{equation} \label {cs1}
\left({d\gamma \over dt}\right)_{cs} =
c \int d\epsilon \int d\Omega n_{ph}(\epsilon, \Omega)
(1 - \beta \cos \Psi)
\int d\epsilon'_s
\int d\Omega'_s
{d\sigma' \over {d\epsilon'_s d\Omega'_s}}
(\epsilon_s - \epsilon)
\end{equation}
where $\epsilon = \hbar \omega / {m_e c^2}$ is the incident photon energy
in units of the electron rest energy, $n_{ph}(\epsilon, \Omega)$ is the
number density of incident background photons within energy and solid-angle
increments $d\epsilon$ and $d\Omega$, and $\Psi$ denotes
the angle between these photons and the local electron beam direction.
The quantity $d\sigma' / d\epsilon'_s d\Omega'_s$ is the magnetic Compton
scattering cross section in the local electron rest frame (ERF), where
the primes denote quantities evaluated in the ERF and the subscript
$s$ labels scattered photon quantities.

In the strong magnetic field the CS cross section includes both nonresonant
and resonant components (Herold 1979, Daugherty and Harding 1986,
Bussard et al. 1986). Dermer (1990) has derived a nonrelativistic
approximation for the loss rate based on the magnetic Thomson cross section
in the ERF (Herold 1979), resolving the total loss rate into component
terms which he labels `angular`, `nonresonant`, and `resonant`.
Sturner (1995) has applied further simplifying assumptions to these terms
in order to derive convenient expressions for the CS loss rate.
His results are summarized in his equations (4)-(9), which we have
incorporated into our acceleration tracing algorithm.

Sturner (1995) notes that for $\gamma \gtrsim 10^3$, the incident thermal
photon energies above the cyclotron resonance may become relativistic
$(\epsilon' \gtrsim 1)$. In this case he replaces the nonresonant component
of the cross section by a relativistic (but nonmagnetic) Klein-Nishina
expression given by his equations (10)-(14). In this work we have
included these expressions, although we note that a more accurate treatment
will require the use of the magnetized (resonant) Klein-Nishina cross section
(Daugherty and Harding 1986, Bussard et al. 1986).

The only loss term which Sturner (1995) includes under the `other` label
in Eq. (\ref {erg}) arises from electron-photon scattering events in which
the scattered photon is replaced by an emergent $e^+/e^-$ pair. Using
cross sections found by Mastichiadis et al. (1986) and Mastichiadis (1991)
for the nonmagnetic form of this process, Sturner (1995) applies a
monoenergetic photon approximation to derive a loss rate given by
his equation (16). For our model parameters this term is never dominant,
but for generality we have also included it in our simulation. As in the case
of his Klein-Nishina CS loss rate, however, we note that in future work the
magnetic form of this process should be investigated since it may also
exhibit resonant behavior which may increase its signifance.
\newpage
\section{SIMULATIONS OF EXTENDED PC CASCADES}

The basic features of our cascade simulation code are described in DH94.
The version used in this work includes several major improvements.
These include revised adaptive algorithms for tracing photon propagation,
which allow more accurate localization of near-threshold pair conversion
events. We have also improved the tracing of synchrotron/cyclotron
emission, which now more accurately simulates both recoil and angular
distribution effects in the cyclotron regime ($\gamma \gtrsim 1$).

However, the most significant improvement for this work is the algorithm
for tracing electron acceleration through extended regions above the PC.
In our current version, each primary electron emerges from the surface
with an initial Lorentz factor $\gamma_0 \gtrsim 1$. Assuming
specific parameters for both the energy gain and loss mechanisms
as described in Section 3 above, the calculation then traces the net
acceleration of the electron
as it escapes outward along the local magnetic field line.
For this purpose we have developed an adaptive numerical technique to
integrate Eq. (\ref {erg}) which accomodates a wide range of
energies and distance scales. To estimate the significance of Compton
losses due to thermal photons from the NS surface we have used a model
similar to that employed by Sturner (1995),
in which the PC has a uniform surface temperature $T_6$ (in units of $10^6 K$)
within a circle of radius $R_{tpc}$ centered on each magnetic pole.
This region is defined as the {\it thermal\/} PC. Note that $R_{tpc}$ may
differ from the PC radius as defined by the locus of the outermost open
field lines. In fact we treat both $T_6$ and $R_{tpc}$ as parameters
in the model. At present we ignore any softer emission which may be
emitted from the overall surface.

Figure 2 shows sample acceleration profiles $\gamma(h)$, where $h$
is the height above the NS surface in stellar radius units.
Curve (a) shows a case in which the accelerating
field is assumed to be constant, namely
$(d\gamma / ds)_{acc} = 5$ ${\rm cm}^{-1}$,
from the surface up to a sharp cutoff at height $h_c = 3$.
Curves (b) and (c) both assume that the gain rate is a linearly
increasing function $(d\gamma / ds)_{acc} = 5h$, over this same region.
They differ only in the assumed values for the Comptonization
parameters, namely the thermal PC temperature $T_6$
and radius $R_{tpc}$ (measured in NS radius units).
Curve (b) assumes $T_6 = 1$ and $R_{tpc} = 0.1$,
corresponding to a cool, small thermal PC. The opposite case of a
hot, large PC, is shown by curve (c) which assumes $T_6 = 2$ and
$R_{tpc} = 0.5$. We note that the constant-acceleration curve (a)
is not affected by these variations of the Comptonization parameters,
since in this case
the gain rate greatly exceeds the loss terms in Eq. (\ref{erg}).
We also observe that even the linear-acceleration profiles are sensitive
to the Comptonization parameters only for heights $h \ll 1$,
and they have little effect on the peak energies reached at the
cutoff height $h_c$.

As $\gamma$ exceeds values $\gtrsim 10^6$ the primary CR emission
reaches gamma ray energies, resulting in photon-pair cascades.
The calculation, as described in DH94, recursively traces the full cascade
development and accumulates 3D tables of emergent $\gamma$-ray counts
vs. energy and solid angle, from which we derive spectra and light curves
of the emission as seen from arbitrary viewing directions.
In this work we have accumulated photon counts from ensembles
of primary electrons distributed in concentric rings over the PC surface.
We have assumed that the primary beam current is
axisymmetric with respect to the magnetic axis, hence the electrons in
each ring are spaced uniformly in azimuth. However, our analysis facility
allows us to assign arbitrary weights to the $\gamma$ counts from each ring.
This technique allows us to vary the assumed radial dependence of the
primary electron current density without requiring new runs of the
simulation.

Finally we should point out that our current simulation is based strictly
on a CR-initiated cascade, i.e. it considers Comptonization as
an energy loss mechanism acting on the primary electrons but it does
not yet include the upscattered photons as a source
of high-energy $\gamma$-rays which may themselves initiate cascades.
This is in obvious contrast to the cascade model proposed by
Sturner and Dermer (1994), in which Comptonization
provides {\it all \/} the high-energy input photons.
Under our model assumptions the primaries reach much higher peak energies
($\gamma \gtrsim 10^6$) than the values they assume ($\gamma \sim 10^5$),
so that in our case CR should initiate the bulk of the cascade emission.
However, we recognize that Comptonization may add a measurable contribution
to the emergent $\gamma$-emission and in a separate work we will extend
the cascade simulation to trace the CS upscattered photons as well.
At the same time, we note that the CS contribution may be expected
to produce a narrower
$\gamma$-beam than the extended CR component we consider here, since it
should originate closer to the PC surface. Thus it is possible that
PC cascades initiated by CR and CS photons may be distinguishable both
spatially and energetically.

\newpage
\section{ELECTRON CURRENTS NEAR THE PC RIM}

In DH94 we showed that single magnetic poles
can exhibit doubly peaked light curves with phase separations
$\delta\phi \lesssim 0.5$ if $\alpha \sim \theta_b$ and the observer
angle $\zeta \sim \alpha$. However, in order
to reproduce the small duty cycles of the double peaks seen in the Crab,
Vela, and Geminga, we had to impose an additional {\it ad hoc\/}
assumption that the primary electron density is strongly concentrated
near the PC rim. We also noted that there are two possibilities for
obtaining doubly peaked profiles with $\delta\phi < 0.5$, in which
the designations of {\it leading\/} and {\it trailing\/} peaks are
reversed. In DH94 we considered in detail the case in which the first
peak corresponds to the phase at which the observer
viewpoint emerges from the interior of the (hollow) $\gamma$-beam, while the
second peak marks the point of reentry. This case, which we denote
as the Exterior Scenario (ES), can produce $\delta\phi < 0.5$ if the
rotational axis is contained within the $\gamma$-beam ($\alpha < \theta_{b}$).
By combining the ES with the assumption that the primary current is
concentrated near the PC rim, we could account for both the short
duty cycles and the lack of emission outside the peaks (since this would
be the phase interval during which the observer viewpoint penetrates
the interior of the hollow beam).
In this scenario we associated the finite emission observed
between the peaks with residual, higher-altitude cascades, which would
produce emission with larger beam widths.

In work following DH94 we have compared our model predictions in detail
with CGRO observations of phase-resolved spectra for the Vela pulsar
(Kanbach et al. 1994). We have concluded that the ES does
not provide uniformly consistent fits to the spectra, especially for
the phase intervals between the main peaks.
In the ES model the high-altitude cascades which produce
the interpeak emission do tend to produce harder spectra below their
characteristic high-energy turnovers, since a smaller fraction of the
hard CR emission is converted to softer cascade photons. By itself
this trend is at least qualitatively consistent with the observations.
However, the peak CR energy ($\propto \gamma^3$) also decreases rapidly
as the primaries lose energy above the acceleration zone, with the result
that the turnovers in the interpeak cascade spectra drop to lower
energies compared to the peak spectra. In this respect the model
prediction is opposite to the observed trend.

This problem with the ES has led us to reexamine the alternative
labeling of the leading and trailing peaks, in which the PC interior
is identified as the source of the interpeak emission. We refer to
this case as the
Interior Scenario (IS). In order to produce finite interpeak emission
in this case, we must abandon the phenomenological DH94 model of a pure
rim distribution for the primary electrons. However, if we replace the
pure rim model with a two-component model which includes a uniform
interior current, it turns out that the IS allows a more consistent
overall agreement with the observations than the ES. Moreover, in this
scenario we can suggest a tentative physical interpretation for a
two-component primary current. In particular, the uniform component
is a simple
approximation of a Goldreich-Julian (GJ) current
$I_{GJ} = \pi {R_{pc}}^2 c \rho_{0}$ (Goldreich and Julian 1969), where
\begin{equation} \label{gjc}
\rho_{0} \sim {{-\bf{\Omega \cdot B}} \over {2 \pi c}}
\end{equation}
which
should be valid if $\theta_{pc} \ll 1$. We propose that this component
includes all the true {\it primary\/} electrons drawn from the
NS surface. In this view the extra rim component consists of secondary
electrons from pairs preferentially created near the PC rim, where the
increasing field-line curvature produces more rapid $\gamma$-pair
conversions.

If any secondary pairs contribute to the PC current of high-energy
particles which initiates cascades, the pairs must themselves
be accelerated to energies comparable with the peak primary energies.
This in turn would require at least some pairs to be created well below
the acceleration cutoff height. If (as we assume here) the primaries
are negative electrons ($e^{-}$), each $e^{-}$ secondary would then move
{\it upward\/} and thus add to the GJ primary current, while the $e^{+}$
would be accelerated {\it downward\/}  along the local field line toward the
surface. In fact the model $\gamma$-ray light curves we present in Section 6
show that if just a small fraction ($\sim 10^{-2}$ or less) of the
cascade pairs created near the rim can be boosted to $\gamma \gtrsim 10^6$,
a two-component current model shows good agreement with observations.

In spite of these results, we must first consider a fundamental
theoretical objection to the acceleration of secondary pairs.
The problem is that the onset of cascade pair production
is expected to produce a sharp cutoff in the acceleration of the
primaries at a height $h_c$, which marks the boundary of the
overlying pair plasma (e.g. Ruderman and Sutherland 1975, Arons 1983).
Our own simulation results confirm that the quenching of
$E_{\parallel}$ above $h_c$ should be an abrupt process, since
the density of created pairs is found to rise sharply with height.
This is demonstrated in Figure 3, which plots typical growth curves of
the multiplicity $M = (N_s^+ + N_s^-)/N_p$ where $N_p$ and $N_s$
denote the numbers of primaries and secondaries
respectively. Thus even if pairs created at the lowest heights can
be accelerated by a decreasing  $E_{\parallel}$ within a finite
transition zone, the growth curves indicate that this zone
is too short for any $e^{-}$ secondaries to reach energies
comparable to those of the primaries. This appears to eliminate the
most obvious model for enhancing the PC current near the rim, in which
the negative pair members are accelerated outward with the primaries.

However, the positron ($e^{+}$) component in such a transition zone
must also be subject to acceleration. The key point here
is that these particles may be drawn {\it downward\/} from the
transition zone back into the acceleration zone, following the local
field lines back toward the NS surface. In fact they should traverse
a distance comparable to the full extent of the acceleration zone,
and thus reach energies sufficient to create (tertiary) pairs by a
variety of possible mechanisms (e.g. $\gamma-B$ pair production,
triplet pair production). The result would be the creation
of pairs deep within the acceleration zone, whose $e^{-}$
members could be accelerated outward with the primaries to reach
similar peak energies.

This sort of cascade feedback process should be most likely to occur
above those regions of the PC where the original upward-directed cascades
initiated by the primaries commence at the lowest heights. Unless the
electrostatic acceleration varies greatly over the PC interior regions,
the increasing curvature of the field lines from the pole to the rim
implies that the primary cascades develop first near the rim (cf. Figure 3).
Hence we argue that reverse $e^+$-acceleration and downward-oriented
cascades occur preferentially around the rim.

As a first test of this hypothesis we have generalized our acceleration
tracing algorithm to follow secondary positrons downward from creation points
just below the cutoff height, back toward the NS surface.
The results confirm that these particles can be boosted to
$\gamma \lesssim 10^7$ at heights $h \gtrsim R_{ns}$ above the surface,
allowing their CR spectra to reach pair-conversion energies and
initiate downward-oriented cascades. In a separate work we will refine
our complete simulation code to investigate the development of these
cascades in detail. We anticipate that their presence may impact our model
in several respects, since in addition to providing a new source of
electrons these cascades can influence the behavior of the acceleration
process just above the surface. In particular, if the cascades
create a sufficiently dense layer of pair plasma overlying the surface
they can retard acceleration below the effective height of this layer.
In addition, it is possible that energetic downstreaming cascade photons
can impose severe Comptonization losses on upward-directed electrons.
As described in the following sections, in this work we will allow for
these possibilities by considering simple models in which the acceleration
may effectively commence at finite heights above the NS surface.
\section{GAMMA-RAY LIGHT CURVES}
The 3D photon count tables accumulated by the simulation may be summed over
energy bins to produce 2D sky maps of the $\gamma$-emission between
arbitrary energy
limits. An example is shown in Figure 4, which plots a grayscale contour map
of emission over 100 MeV. Any horizontal line drawn across this plot
corresponds to a specific value of the polar angle $\zeta$ for a given
viewing direction, and the counts distributed along this line define
the $\gamma$-ray light curve as seen from this viewpoint.

Following the arguments in Section 5 we present sample results for
the Vela pulsar using a simple two-component primary current model,
which we obtain by superimposing simulation datasets for concentric
rings of primaries as discussed in Section 4. In each case we have
included a total of 10 rings spaced at equal radial increments
to cover the PC interior. Since each ring contains 180 particles
with a uniform 2-degree azimuthal spacing, the inner rings are
weighted $\propto r^{-1}$ to approximate a uniform interior density.
To simulate test cases with a moderate rim component, we have weighted
the outermost ring by arbitrary factors in the range 3 to 5. Physically
this corresponds to the acceleration of a few secondary electrons for
each primary electron on this ring, which is a small fraction of the
$10^{2}-10^{3}$ cascade pairs created per primary near the rim.

All the datasets we have accumulated to date assume the following
general form for the accelerating field, namely
\begin{equation} \label{acc1}
E_{\parallel}(h) = {{m c^2} \over e} \left({d\gamma \over ds}\right)_{acc}
= {{m c^2} \over e} [a_0 + a_1 (h - h_0)] \Theta(h - h_0) \Theta(h_c - h)
\end{equation}
We choose units for Eq. (\ref{acc1}) such that the path length $s$ is
measured in cm, while the height $h = (R - R_{ns})/R_{ns}$ is in NS
radius units from the PC surface, $\Theta (x)$
is the unit step function ($0$ for $x < 0$, $1$ for $x > 0$),
and the constants $a_0$, $a_1$, and $h_0$ are taken as free parameters
in our model. Their values effectively determine the height at which
cascades commence above the PC rim, which Arons (1983) denotes as the
``pair formation front''. In the following we take the height at
which the cascade multiplicity exceeds unity (cf. Figure 3)
as a reasonable measure of the acceleration cutoff height $h_c$.
Thus $h_c$ is a function of $(a_0, a_1, h_0)$ but is {\it not\/}
itself a free parameter. In practice we determine $h_c$ from trial
simulations before generating complete datasets.

The quantity $h_0 \ge 0$ in Eq. (\ref{acc1}) denotes the height at which
acceleration commences. We introduce $h_0$ to allow for the possibility
that downward-oriented cascades may prevent or impede acceleration just
above the NS surface. As noted in Section 5, this can occur
either if the cascades create a sufficiently dense layer of pair plasma
overlying the surface, or if downstreaming cascade photons impose strong
Comptonization losses on upward-moving electrons. In a separate study
of downward-oriented cascades we will investigate both of these effects
in order to put physical constraints on the choice of $h_0$, but here
we simply explore the effects of varying $h_0$ in sample models.

If we let $a_1 = 0$ in Eq. (\ref{acc1}) we obtain a constant-field
approximation, resembling vacuum gap acceleration models of the type
proposed by Ruderman and Sutherland (1975). If instead we set
$a_0 = 0$, we have a crude approximation for the potential suggested by
Arons (1983) in his slot-gap model. We have generated datasets for the
Vela pulsar using both of
these limiting forms. In each case we have empirically chosen combinations
of the parameters $(a_0,a_1,h_0)$ such that the primary electrons
reach their peak energies ($\gamma \gtrsim 10^6$) rapidly enough to
initiate cascades. We note that in this work we have assumed no dependence
of either $a_0$, $a_1$, or $h_0$ on the magnetic polar angle $\theta$.
We have used a further simplifying assumption here, namely that
the cutoff height $h_c$ has the same value over the PC interior
as determined by the onset of cascades near the rim. While this assumption
must be questioned in a more refined treatment, we show below that it
leads to encouraging agreement with observations.

Figure 5(a,b,c,d) shows model light curves obtained under these assumptions
for the acceleration function Eq. (\ref{acc1}), using sample parameters
$(a_0,a_1,h_0) = (5,0,0)$, $(0,5,0)$, $(0,20,1)$, and $(50,0,2)$,
which we denote as models A, B, C, and D respectively.
Table 1 lists additional simulation parameters which are common to all
these models. In models A-C a common weight factor of 5
was assigned to the outermost ring of primary electrons to represent the
excess rim current, while a factor 3 was used for model D. (The simulation
would assign a weight factor of 1 to this ring for a uniform distribution.)
In each case the rim weights were chosen to obtain reasonable fits to the
observed Vela light curve. Since each simulation includes a total of 10
concentric primary current rings covering the PC interior, these rim
weight factors increase the total PC currents above their uniform component
values by factors of roughly 1.7 for models A-C and 1.4 for model D.
For comparison, in each of these plots the light curves which would be
produced by the uniform current alone (without the excess rim component)
are shown in gray.

If we compare these model results with the observed Vela
light curve (Kanbach et. al 1994) shown in Figure 9, we see that
the acceleration parameters which best match the observations are those
for which the acceleration near the surface is low. In fact, satisfactory
fits are obtained only if the
primaries do not reach $\gamma \gtrsim 10^6$ until after they have
attained heights $h \gtrsim 1$. If they exceed these energies at altitudes
too far below the cutoff height $h_c$, the total cascade emission which
they produce over the full acceleration region and beyond is spread
over large solid angles, yielding broad pulse peaks.
In particular, this tendency rules out constant-acceleration models
($a_1 = 0$) such as that shown in Figure 5(a), except in cases where
$h_0 \gtrsim 2$ as in Figure 5(d). A comparison of Figures 5(b), 5(c)
shows that even for linear acceleration ($a_0 = 0$), the fits are
significantly improved by introducing nonzero values of $h_0$.

Among the sample runs shown in Figure 5, models C and D show peak duty
cycles which are in the best agreement with the observed values. Moreover,
in each of these cases the first half of the interpeak emission
resembles both the magnitude and slope seen in the data. This example
shows that the two-component model for the primary current can produce
consistent agreement with a significant portion of the total light curve.
Unfortunately the agreement breaks down for the
trailing interpeak component, but since our model assumes axisymmetric
current rings it cannot account for any strong asymmetry in the light curve.

Finally we note that all these models predict a low but finite level of
emission
throughout the phase interval between Peak 2 and Peak 1 (i.e., over regions
outside the PC rim). This emission
is due to the residual, high-altitude cascades which we suggested in DH94
might be the source of the interpeak emission. Kanbach et al.
(1994) find no detectable emission in this phase interval for Vela,
and no evidence for unpulsed emission. Given their stated estimates
for the EGRET detector sensitivity, however, their findings are not in
conflict with our model results for the sample datasets C and D described
above. However, the observations do impose an additional constraint on the
relative weight factors for the two-component PC current distribution.
For example a uniform PC current, without any rim current enhancement,
would produce significantly more emission outside the peaks than the
observations allow.

\section{PHASE-RESOLVED ENERGY SPECTRA}
The same choices of parameters (model C and D) which best match the
observed light curves in Section 6 also produce the best fits for the
energy spectra. In spite of the similar appearance of their light curves,
however, model D produces better spectral fits than model C. In fact,
as shown in Figure 6 model D provides the closest match to the observed
total (phase-averaged) spectrum across five decades in energy,
The spectral differences among these models
are principally due to their varying extent of cascade development.
In models A and B the primary electrons reach maximum energies of
$7.5 \times 10^6$ and $1.2 \times 10^7$ respectively, compared to
$1.7 \times 10^7$ for Model C and $2.0 \times 10^7$ for Model D.
The values for models A and B especially are too low to supply
either the photons up to 3 GeV or the level of emission observed below
100 MeV.  Clearly, the observed Vela total emission is not the result
of curvature radiation alone.

Figure 7 shows that this agreement for model D applies not only to the
total spectra, but also to the phase-resolved spectra observed by EGRET
(Kanbach et. al. 1994). These plots show fits for various phase intervals
defined by these authors in their power law-fits to the Vela
phase-resolved spectra for energies between 70 and 4000 MeV.
The normalization factors were determined separately at each phase interval
to match the data and differ by less than a factor of 2.
We note that the model
reproduces the tendency for the (quasi) power-law spectra at the phase
intervals of the two peaks to become significantly softer than the spectra for
the interpeak subintervals. In the Interior Scenario (Section 5)
this trend is expected since the interpeak emission is due to the interior
primary electron current, whose hard CR emission is less efficiently
converted to softer cascade photons (cf. Figure 3). The IS model also
reproduces the
observational feature that the high-energy turnovers in the Vela spectra
occur at lower energies for the peaks vs. the interpeaks.  The sharpness of
the high-energy turnovers in the P1 and P2 spectral intervals, due to
magnetic one-photon pair production attenuation, are also reproduced,
especially in P1.
The model D spectra in the phase intervals LW1 and TW2, the emission just
outside the peaks, turnover more gradually and at energies below 500 MeV.
This emission is primarily curvature radiation at high altitudes from primary
electrons that have lost a significant amount of their maximum energy.
These phase intervals are thus predicted to have the softest spectra,
consistent with both the data and the high indices of the power law
fits of Kanbach et al. (1994).  In model D, the hard spectra in
intervals I1 and I2 extend to energies below 10 MeV, predicting that the
interpeak emission should decrease relative to that of the peak emission
at lower energies.  This appears to be verified by the 0.07 - 0.6 MeV
light curves measured by OSSE (Strickman et al. 1995), where no interpeak
emission was detected.

One quantitative measure of the spectral evolution during each pulse
is the hardness ratio $H$, defined here as the ratio of the flux over
300 MeV to the flux between 100 and 300 MeV. Figure 8 shows the model
D hardness ratio vs. pulse phase for $\zeta = 16^{\deg}$, corresponding
to the phase-resolved spectra in Figure 7. The trend toward harder spectra
during the interpeak phase interval is clear and appears to be consistent
with EGRET Vela observations (Fierro et al. 1995).
\newpage
\section{TOTAL GAMMA FLUX ESTIMATES}
If we identify the uniform component of our model PC current with the
GJ current predicted by Eq. (\ref {gjc}), we can estimate an upper limit on
the absolute $\gamma$-ray flux levels expected from our model sources
within any specified energy range $\Delta E_\gamma$. The required inputs
are the dataset sky map counts, the pulse period $P$, and the estimated
distance $D$ (which we take to be 500 pc for Vela). We outline the procedure
briefly as follows.

First we derive the effective number of primaries traced in the simulation,
taking into account the weight factors assigned to each concentric ring
of electrons. Following the arguments in Section 5, we resolve this total
number of primaries into two components representing uniform and rim
distributions respectively. As noted above, our total flux estimate
(uniform plus rim components) assumes that the uniform component is
a GJ current. For the flux estimate, the quantity of interest is the
number of GJ primaries in the simulation.

After summing the full 3D photon arrays over the energy range
$\Delta E_{\gamma}$ to produce the appropriate 2D sky maps, we find
the number $\Delta N_\gamma$ of photons accumulated along a 1-bin strip
of constant $\zeta$ and angular width $d \zeta$ during one full pulse
($\Delta \phi = 2 \pi$). The phase-averaged $\gamma$-ray flux $F_\gamma$
{\it per primary electron\/} at the distance D is then given by
\begin{equation} \label{flx}
F_\gamma = \Delta N_\gamma / 2 \pi \sin \zeta d \zeta P D^2 N_{GJ}
\end{equation}
where $N_{GJ}$ denotes the effective number of GJ primaries in the
dataset (excluding the excess rim component). Finally we obtain an absolute
total flux estimate by multiplying Eq. (\ref{flx}) by the (maximal) current
of GJ primaries from the PC surface as given by Eq. (\ref{gjc}).

The predicted fluxes for our Vela models A,B,C,D at energies $ > 100$ MeV
as found from this procedure are
$7.3 \times 10^{-5}$,
$1.6 \times 10^{-4}$,
$2.8 \times 10^{-4}$,
$2.8 \times 10^{-4}$
photons ${\rm cm}^{-2} {\rm s}^{-1}$ respectively.
It turns out that these values are all an order of magnitude
higher than the average flux observed by EGRET (Kanbach et al. 1994),
namely $(7.8\pm 1.0) \times 10^{-6}$ photons ${\rm cm}^{-2} {\rm s}^{-1}$
for $E_{\gamma} > 100$ MeV. Our high model flux levels, which obviously
are due to strong beaming factors of the hollow-cone emission,
are not by themselves a problem for our model since the GJ estimate
should properly be regarded only as an upper limit on the PC current.
We note, however, that the model flux estimate does fall closer to the
GJ limit as the $\gamma$-beam half-angle $\theta_b$ is increased.
In this respect
the excess predicted flux shows that even larger PC dimensions and/or
acceleration cutoff heights can be allowed within the framework of
the model.

\section{COMPARISON WITH OBSERVATIONS AT OTHER WAVELENGTHS}
In the preceding sections we have applied the SPC model specifically to
the Vela pulsar, in part because both the $\gamma$-ray light curves and
phase-resolved spectra for this object have been observed in
considerable detail. However, our model results for Vela can also
account in general terms for the $\gamma$-ray emission from other
pulsars with doubly-peaked profiles such as the Crab, Geminga, and
PSR B1951+32 (Ramanamurthy et al. 1995). The second general class
of light curves predicted by the SPC model, namely those with only
a single broad peak, may describe PSR B1055-52 (but see below).
At present the only source whose $\gamma$-ray light curve may be
difficult to accomodate is PSR B1706-44 (Thompson et al. 1992), since
recent EGRET observations (Thompson et al. 1995) suggest that this object
may have a triply-peaked pulse.

However, we must consider whether the SPC $\gamma$-ray model is
also compatible with observations of pulsed emission at other wavelengths
from Vela and the other known $\gamma$-ray pulsars. Our primary
concern here involves the possible implications of these observations
regarding the viewing geometry for each source. In this context we focus
especially on three $\gamma$-ray pulsars for which we also have strong
evidence of thermal X-ray emission from the NS surface, namely
Vela itself (\"Ogelman et al. 1993), Geminga (Halpern and Holt 1993),
and PSR B1055-52 (\"Ogelman and Finley 1993).
These objects are of particular interest since the modulation and phase
behavior of the X-ray emission should be directly related to the magnetic
field geometry at the NS surface.
To facilitate the discussion of these sources, in Figure 9 we have assembled
their light curves at various wavelengths using a common phase origin for
each source. It turns out that each object presents a distinct set of
challenges for our model, which we analyze separately below.

Although it shows no evidence of surface thermal X-ray emission we must
also consider observations at other wavelengths from the Crab pulsar.
The Crab has the distinction of having doubly peaked light curves in
phase at all observed wavelengths. However, its optical emission exhibits
polarization swings which cause special problems for the SPC model.
An additional challenge is presented by recent HST and ROSAT imaging
of the inner Crab nebula, which strongly suggest an observer
angle $\zeta \lesssim 60^{\deg}$ (Hester et al. 1995).

(a) Vela (PSR B0833-45)

As shown in Figure 9(a), the pulsed radio emission from Vela
(see for example Manchester and Taylor 1977) exhibits a single narrow peak
which leads the first $\gamma$-ray peak by
$\sim 0.12$ in phase (Kanbach et al. 1994). The radio pulse shows a high
degree of linear polarization with an unusually wide
swing ($\gtrsim 90^{\deg}$) in the polarization angle $\psi$ across the
pulse. This behavior has been interpreted (Radhakrishnan and Cooke 1969,
see also Michel 1991) in terms of the rotating projection of a dipolar
magnetic field in the plane orthogonal to the viewing
direction. In this model $\psi$ is given as a function of $\alpha$,
$\zeta$, and the pulse phase angle $\phi$ by
\begin{equation} \label{pol}
\tan \psi = \sin \alpha \sin \phi
  / (\sin \zeta \cos \alpha - \cos \zeta \sin \alpha \cos \phi)
\end{equation}
Several authors (e.g. Lyne and Manchester 1988, Rankin 1990) have attempted
to invert this relation to determine the values of $\alpha$ and $\zeta$ for
various pulsars, although the results to date are subject to controversy
(Michel 1991, Miller and Hamilton 1993). However, Eq. (\ref{pol}) does
imply that the maximum rate of the polarization
swing $R \equiv |d(\tan \psi) / d(\sin \phi)|$ occurs at the phase
corresponding to the closest approach of the magnetic axis to the observer
direction, which we denote by $\phi_{M}$.
If this model is correct, the rapid, extended swing for Vela
($R \sim 5.9$) indicates that the observer viewpoint approaches a
magnetic pole to within a few degrees.

As may be seen from Figure 10, the values of $\alpha$ and $\zeta$ used
in the Vela model datasets discussed in Sections 6 and 7
do not produce polarization swings which are either as rapid or extended
as the observed values. However, the real challenge in accounting
for the radio pulse in our model is not simply to find better combinations
of these parameters. The key point is that if the radio pulse does indeed
mark the phase of closest approach to either of the magnetic poles,
in the case of Vela its location relative to the $\gamma$-ray peaks is
inconsistent with the SPC model. In particular, the Interior Scenario
requires $\phi_{M}$ to lie midway between the two $\gamma$ peaks, whereas
in the Exterior Scenario it is displaced from the midpoint by $0.5$
in phase. In contrast, Kanbach et al. (1994) find the phases of the
$\gamma$ peaks (relative to the phase of the radio peak, $\phi_0 = 0$)
to be
$\phi_{p1} = 0.12$ and $\phi_{p2} = 0.54$ respectively.
Hence the standard PC model of the radio pulse asserts that
$\phi_{M} = \phi_0 = 0$, while the IS predicts
$\phi_{M} = (\phi_{p1} + \phi_{p2})/2 = 0.33$ and the ES has
$\phi_{M} = 0.83$. Thus the standard model of the Vela radio pulse
is inconsistent with the SPC $\gamma$-ray model.

On the other hand, it turns out that both the optical and X-ray light curves
for Vela fit much more naturally within the geometry of the IS.
As shown in Figure 9(a), the optical emission (Wallace et al. 1977)
has a doubly peaked light curve with a smaller peak-to-peak phase separation
($\sim 0.2$) than that seen in the $\gamma$-ray regime. Moreover, the
$\gamma$-ray peaks enclose the optical
peaks in the sense that the leading optical peak follows the leading $\gamma$
peak, while the opposite occurs for the trailing peaks (see for example
Manchester and Taylor 1977). In the IS, this sort of optical/$\gamma$ phase
relationship would hold if the optical and $\gamma$ emission were beamed in
coaxial hollow cones from the PC, with beam angles
$\theta^{opt}_{b} < \theta^{\gamma}_{b}$. This in turn suggests
that the optical emission might either be associated with interior
PC currents, or that it might be produced by the rim current at lower
heights than the $\gamma$-emission.

Figure 9(a) also shows the pulsed X-ray emission from Vela detected
by the ROSAT satellite (\"Ogelman et al. 1993), which consists of a
broad pulse trailing
the radio peak, with the bulk of the emission
occurring between the two $\gamma$-ray peaks. The harmonic content of the
pulse suggests a complex nonsinusoidal structure, although the available
X-ray data do not show firm correlations with the optical or $\gamma$
peaks (or clear evidence of more than one peak). The statistics
are unfortunately limited by the fact that the emission contains
contributions from the compact nebula as well as the pulsar,
and the pulsed fraction of the latter is only about 11\%.
\"Ogelman et al. (1993) obtain their best fit to the pulsed component
with a soft blackbody spectrum ($T_6 \sim 1.5-1.6$). They also note that
the total point source (pulsed plus unpulsed) can either be fit with a
blackbody spectrum at a similar temperature or with a steep power law
($\Gamma \sim -3.3$), compared to a harder power law ($\Gamma \sim 2.0$)
which fits the surrounding compact nebula. \"Ogelman et al. (1993)
suggest that if the pulsed component is actually thermal emission,
the modulation may be due either to a nonuniform surface
temperature distribution or to anisotropic radiation transfer
effects in the magnetosphere. In either case the key point
for our model is that the pulsed X-ray emission should then be concentrated
near the phase $\phi_{M}$ of closest approach of the observer direction to
a magnetic pole (Page 1995). To the extent that the bulk of the emission
does occur between the $\gamma$ peaks, the Vela X-ray light curve appears
compatible with the IS $\gamma$-ray model.

In summary it appears that the observed optical, X-ray, and $\gamma$-ray
light curves for Vela all seem mutually consistent with the IS,
whereas the radio polarization swing cannot have the usual interpretation
based on Eq. (\ref{pol}) in either the IS or the ES.
At present we have no satisfactory way to account for the phase of the
Vela radio pulse within
the general framework of any SPC model, unless we invoke the
possibility of nondipolar magnetic fields near the NS surface.

However, we should point out that this incompatiblity is not simply a
problem for our $\gamma$-ray model. The same conflict already exists
between the standard radio model and the entire class of thermal X-ray
models (e.g. Page 1995) in which the peak(s) in the pulsed emission
coincide with the closest approach of the magnetic pole(s) to the
observer viewpoint.

(b) Geminga (PSR B0630+178)

Although Geminga has long been known to be a strong $\gamma$-ray source
(Kniffen et al. 1975), it was first discovered to be a pulsar from
X-ray observations (Halpern and Holt 1992). Shortly thereafter
$\gamma$-ray pulses were detected at the X-ray period (Bertsch et al. 1992).
To date no pulsed emission has been found at either radio or optical
wavelengths, although an optical counterpart has been identified
(Bignami et al. 1993).

While the lack of optical and radio light curves prevent the sort of
phase comparisons we can make for other sources, both the X-ray and
$\gamma$-ray data are relatively rich in detail. Figure 9(b) shows
the light curves for Geminga at both hard and soft X-ray energies
from ROSAT observations (Halpern and Ruderman 1993) as well as in
the EGRET $\gamma$-ray regime (Mayer-Hasselwander et al. 1994,
Ramanamurthy 1995).
As in the case of Vela, the $\gamma$-ray light curve above 100 MeV
exhibits a two-peak
structure with significant interpeak (bridge) emission. The peaks have
duty cycles only moderately larger than in Vela, with a phase separation
of $0.5$. In contrast, Halpern and Ruderman (1993) find
that the X-ray light curves at both soft (0.07-0.53 keV) and hard
(0.53-1.50 keV) energies consist of broad single pulses. The hard
component is somewhat narrower, but perhaps most remarkably the soft
and hard components are $\sim 105^{\deg}$ out of phase.

Halpern and Ruderman (1993) have fit the hard and soft components of the
pulsed X-ray spectrum to two blackbody sources at temperatures
$T_{6} \sim 0.5$ and $\sim 3$ respectively. These authors suggest
that the soft emission is from the overall NS surface, while the hard
component arises from hotter regions around a PC. However, they also
note that within the available statistics a power-law fit for the harder
component is nearly as good as the blackbody fit, which leaves open
the possibility of magnetospheric emission mechanisms. In any event
the hot PC model of the hard X-ray emission appears to be consistent
with the SPC $\gamma$-ray model, as in the case of Vela, since as seen
in Figure 9(b) the bulk of the hard X-ray pulse from Geminga also lies
between the double $\gamma$-ray peaks (Halpern and Ruderman 1993).

Unfortunately the modulation of the soft X-ray component and its phase
shift relative to the hard component complicate this model.
In fact the hard and soft components may not be consistently
explained within the framework of {\it any\/}
NS heating/cooling models which assume dipolar magnetic-field symmetry.
This point has led Halpern and Ruderman (1993) to suggest
an off-axis dipole model in the case of Geminga.

(c) PSR B1055-52

This source has been detected by EGRET at energies above 300 MeV
(Fierro et al. 1993). Figure 9(c) shows that in contrast to the doubly
peaked radio pulse, the $\gamma$-ray light curve appears to exhibit
a single broad peak. However, the available statistics are insufficient
to rule out a multipeaked substructure. The limited data makes it difficult
to analyze the phase relationship between the radio and $\gamma$ pulses,
although it may be significant that the precursor of the main radio pulse
appears just at the trailing end of the $\gamma$ peak. It is noteworthy
that the radio profile has some similarity to that of the Crab,
including a peak-to-peak phase separation $\gtrsim 0.4$ which would
require an off-axis dipole in an orthogonal rotator model.

PSR 1055-52 has the distinction of exhibiting the hardest phase-averaged
$\gamma$-spectrum of all the $\gamma$-ray pulsars known to date,
with a photon spectrum index of $\sim 1.2$. It is worth noting here that
PC cascades can definitely exhibit such hard spectra, although they tend
to do so only when both the electron CR losses and pair-conversion rates are
comparatively low. These conditions are most likely to apply in specific
regions of the magnetosphere, especially close to the magnetic axes and/or
at heights of several NS radii above the surface. However, both more
detailed $\gamma$-ray observations and further modeling of this source
will be required to determine how the hardness of the spectrum may
constrain the SPC model.

Pulsed X-rays have also been detected from PSR B1055-52
by ROSAT (\"Ogelman and Finley 1993). As in the case of Geminga, the emission
exhibits distinct hard and soft components above and below $\sim 0.5$ keV,
both of which exhibit broad single pulses. Figure 9(c) shows the phase
relationships between the X-ray light curves and the pulses at
radio and $\gamma$-ray energies. As in the $\gamma$-ray regime, evidence
for substructure in either X-ray component is limited by the available
statistics. Another striking similarity with Geminga is the large
relative phase shift between the hard and soft X-ray peaks, with the hard
component in this case leading
by $\sim 120^{\deg}$. \"Ogelman and Finley (1993) obtain satisfactory
spectral fits using two-component blackbody models,
although they find that the hard component may also be fit by a power law
which extrapolates up to flux levels in the $\gamma$-ray regime
comparable with the EGRET observations.

If PSR B1055-52 does in fact have only one $\gamma$-ray peak, then its
relationship to the X-ray emission may be difficult to explain within
the SPC model. The key problem is that the model identifies the phase
of a single $\gamma$ peak with the phase $\phi_M$ of closest approach
of the PC.
However, if the hard X-ray component is due to PC heating as proposed
for Geminga (Halpern and Ruderman 1993), the X-ray peak indicates a value
for $\phi_M$ in apparent conflict with the $\gamma$-ray location.
While this difficulty does not arise if the hard X-rays have a
magnetospheric origin as \"Ogelman and Finley (1993) suggest, their phase
shift relative to the $\gamma$-ray pulse is still problematical.

As in the case of Geminga, however, the modulation of the soft X-ray
component and its phase shift relative to the hard component
complicate the picture. The fact that the radio pulse for PSR B1055-52
has two peaks, with noteworthy similarities to the Crab radio profile,
is also puzzling. However, the principal question regarding
the viability of the SPC model for this source is whether the
$\gamma$-ray light curve is singly peaked. Hopefully further analysis
of EGRET data will be able to resolve this question.

(d) The Crab Pulsar (PSR B0531+21)

In constrast to all other $\gamma$-ray pulsars, the light curve of the
Crab exhibits a doubly peaked structure at at all wavelengths observed
to date, with the peaks appearing at essentially the same phase positions
throughout the entire spectrum. In purely geometric terms this phase
synchronization seems to suggest that a variety of emission processes,
which may occur in distinct magnetospheric regions of other pulsars,
are spatially coincident in the Crab. In the context of SPC $\gamma$-ray
models it appears to motivate a search for radio, optical, and
X-ray emission mechanisms involving the cascade pairs.

Unfortunately, this approach leads to at least one serious difficulty
for the SPC model, namely the optical polarization swings found to occur
across each peak (Smith et al. 1988). If both the optical and $\gamma$
peaks do originate from the same PC rim regions, then the optical
swings cannot be due to the sort of rotational projection effect
described by Eq. (\ref{pol}) since the extent of the swing
through the phase intervals containing each $\gamma$-peak cannot exceed
a few degrees (cf. Figure 10). However, SPC models for the Crab appear
to be compatible in this respect with the radio pulses, which do not
exhibit significant polarization swings.

In addition to this problem, a significant constraint on SPC models of
the Crab pulsar is posed by recent HST and ROSAT observations of the inner
nebula (Hester et al. 1995). These observations appear to confirm
numerous earlier
suggestions that the observer angle $\zeta$ for the Crab is considerably
larger ($\lesssim 60\deg$) than the values ($\sim 15\deg$) used in our
sample Vela datasets. However, this finding does not by itself rule out
the SPC model for the Crab, since it turns out that such large values of
$\zeta$ can be accomodated if we allow the PC dimensions to be
$\sim 4 \-- 5$ times larger than the standard estimate Eq. (\ref{tpc}),
as opposed to the factor $2$ used in our model datasets for Vela.
Somewhat smaller values are also adequate if the cascades are assumed
to extend up to heights $\gtrsim 3$ NS radii. Thus in the case of the
Crab especially, the dimensions of the PC are critical to our model.
\newpage
\section{DISCUSSION}
The model we have presented here has at least one significant advantage
over an alternative SPC model (Sturner and Dermer 1994, Sturner et al. 1995)
in which PC cascades are initiated by Comptonizaton of primaries by soft
photons from the NS surface rather than CR emission. As we have shown,
extended primary acceleration can easily generate CR-induced cascades
at heights reaching up to several NS radii. In contrast, cascades due
to Comptonization should be confined to significantly lower regions
unless some mechanism for strong beaming of the soft photons is invoked.
Assuming that similar PC dimensions are used in both models, the
Comptonization model has a more limited ability to overcome the
observability problem.

The best results we have obtained to date from the extended cascade SPC model
are for those cases in which the net electron acceleration becomes significant
only at heights $h \gtrsim R_{ns}$ above the NS surface.
However, we have shown in Section 3 that neither resonant Compton scattering
of thermal photons from the NS surface nor other known energy loss processes
considered in previous PC models can effectively counteract
accelerating potentials of the types we have considered over distances
of this order. This applies in particular to resonant Compton scattering,
even if we assume the highest plausible values for both the surface
temperatures and thermal PC radii. Thus it is obviously important to
investigate the possibilility noted in Section 5, namely that
downward-oriented cascades initiated by reversed secondary
acceleration can prevent or impede acceleration just above the surface.
An obvious next step in the exploration of the SPC
model is to trace the development of downward-oriented cascades in detail,
and if possible to estimate both their significance as a source of
energetic Comptonizing photons and the depth of the surface plasma layer
which they may create.

The discussion in Section 9 shows that the phase relationships between
light curves at different wavelengths are in fact quite complex.
The problem of accounting for all these observations in a self-consistent
manner may eventually force us to consider models with asymmetric magnetic
field geometries. One initial step in this direction would be to consider
off-axis dipolar models of the type suggested by Halpern and Ruderman (1993)
in more detail. In such models we anticipate that the modulation of
thermal X-ray emission from, say, the PC surface may be significantly
out of phase with magnetospheric emission produced above the surface
and directed along the open field lines.
\section{ACKNOWLEDGEMENTS}
We are indebted to Joe Fierro, Gottfried Kanbach, Peter Michelson,
P.V. Ramanamurthy, and David Thompson for valuable discussions regarding
EGRET observations, and to Mark Strickman for information regarding OSSE
and COMPTEL results. We also thank Hakki \"Ogelman and John Finley for
discussions on the pulsed X-ray emission from Vela. Michal Marko provided
valuable assistance in the development of our visualization and analysis
software. We gratefully acknowledge support for this work from NASA CGRO
Guest Investigator Grants for Phases 3 and 4 (JKD, AKH), and from the
NASA Astrophysics Theory Program (AKH).

\newpage
\centerline{TABLE 1: Cascade Vela Model Parameters}
\begin{tabular}{ll}
\hline \\
 Period   &                        $P = 89$ ms \\
 Surface Magnetic Field    &       $B = 3 \times 10^{12}$ Gauss \\
 Inclination  &                    $\alpha = 10^{\deg}$ \\
 NS Radius   &                     $R_{ns} = 10^6$ cm \\
 PC Radius   &                     $R_{pc} = 2 R_{ns}
                      \sin^{-1} \left({R_{ns} \Omega / c}\right)^{1/2}$ \\
 PC Surface Temperature  &         $T_{pc} = 2 \times 10^{6} K$ \\
 Thermal PC Radius  &              $R_{tpc} = 0.5 R_{ns}$ \\
 Initial Primary Lorentz Factor &  $\gamma_0 = 1.0$ \\
\\
\hline \\
 Acceleration parameters $(a_0, a_1, h_0)$: \\
 Model A:  & $(5,  0, 0)$ \\
 Model B:  & $(0,  5, 0)$ \\
 Model C:  & $(0, 20, 1)$ \\
 Model D:  & $(50, 0, 2)$ \\
\\
\hline
\end{tabular}

\newpage
\begin{figure}
\centerline{FIGURE CAPTIONS}
\caption{Angle $\psi$ between magnetic axis and tangent to fixed
      dipole field line $r = k \sin^2 \theta$, as function of radial
      distance $r$ from NS center. Curves labeled $1, 2, ...$ correspond
      to field lines originating from NS surface at polar angles
      $\theta_{pc}, 2 \theta_{pc}, ...$, where  $\theta_{pc}$ is PC
      half-angle estimated by Eq. (1).
}

\caption{Lorentz factor $\gamma$ vs.
      height $h$ in NS radius units, measured from NS surface.
      Curve (a) assumes constant-field acceleration model
      $(d\gamma/ds)_{acc} = 5$ for $0 < h < 3$, while (b) and (c)
      both assume linear form $(d\gamma/ds)_{acc} = 5h$ over same region
      for distinct combinations of Comptonization parameters.
      (b) $T_6 = 1$, $R_{tpc} = 0.1$; (c) $T_6 = 2$, $R_{tpc} = 0.5$.
}

\caption{Multiplicity $M$ (number of secondary electrons produced per
      primary electron) vs. radial distance from NS center. Separate growth
      curves are shown for each primary electron ring in sample Vela dataset
      defined in Table 1. These plots assume that
      $E_{\parallel} \propto a_1 h$
      for $h > 0$, where $a_1$ is independent of magnetic colatitude $\theta$
      over the PC). However, acceleration is assumed to cut off abruptly
      above height $h_c$ at which first pairs appear, which is a decreasing
      function of $\theta$.
}

\caption{Angular intensity distribution of gamma emission above 100 MeV,
      plotted using a linear 10-level grayscale.
}

\caption{(a,b,c,d) Simulated Vela $\gamma$-ray light curves for emission
      above 100 MeV, using Model A,B,C,D parameters respectively
      (see Table 1). Corresponding light curves due to uniform PC currents
      alone (neglecting rim components) are shown in gray.
      In each case, observer angle $\zeta$ has been chosen to produce
      peak-to-peak phase separations closest to observed value of $0.424$.
      $\zeta = 12^{\deg}$ for Model A, $15^{\deg}$ for Models B and C,
      $16^{\deg}$ for Model D.
}
\end{figure}
\newpage
\begin{figure}
\caption{(a,b,c,d) Total pulsed energy spectra, $E^{2} {dN / dE}$, for Vela
      Models A,B,C,D and same observer angles $\zeta$ as in Figure 5. Solid
      lines show emergent cascade gamma emission, while dashed lines show pure
      CR emission (ignoring magnetic pair production and cascade formation).
      Data points show observations by EGRET (Kanbach et al. 1994)
      as well as COMPTEL and OSSE (Strickman et al. 1995).}

\caption{Vela Model D phase-resolved spectra, $E^{2} {dN / dE}$, for observer
      angle $\zeta = 16^{\deg}$ and phase intervals used in analysis of
      EGRET data (Kanbach et al. 1994). Solid line shows cascade $\gamma$
      emission, while dashed line shows pure CR emission.}

\caption{Phase-resolved hardness ratios for Vela Model D, assuming
      $\zeta = 16^{\deg}$ as in Figure 7.}

\caption{Relative phases of X-ray and $\gamma$-ray pulses for $\gamma$-ray
      pulsars which appear to emit thermal X-ray emission from the NS
      surface. Where applicable, emission at radio and optical wavelengths
      is also shown. (a) Vela; (b) Geminga; (c) PSR B1055-52.}

\caption{Swing of radio linear polarization angle $\psi$ predicted by
      Eq. (11) for parameters $\alpha = 10^{\deg}$,
      $\zeta = 16^{\deg}$. Dashed lines
      show phase intervals
      containing $\gamma$-ray peaks in light curves for these models.
      Maximum predicted slope of polarization curve for these choices
      of $(\alpha,\zeta)$ is $|R| \sim 1.7$, compared to observed
      value of $\sim 6.5$. Discrepancy in slope indicates that actual
      observer viewpoint has closer approach to magnetic pole than
      model values $(\alpha,\zeta)$ allow. Phase location of radio peak
      poses more serious problem, since dipolar versions of both SPC
      $\gamma$-ray model and thermal surface emission models of pulsed
      X-ray emission suggest that phase of closest PC approach is
      incompatible with standard PC model of radio polarization swing.}
\end{figure}
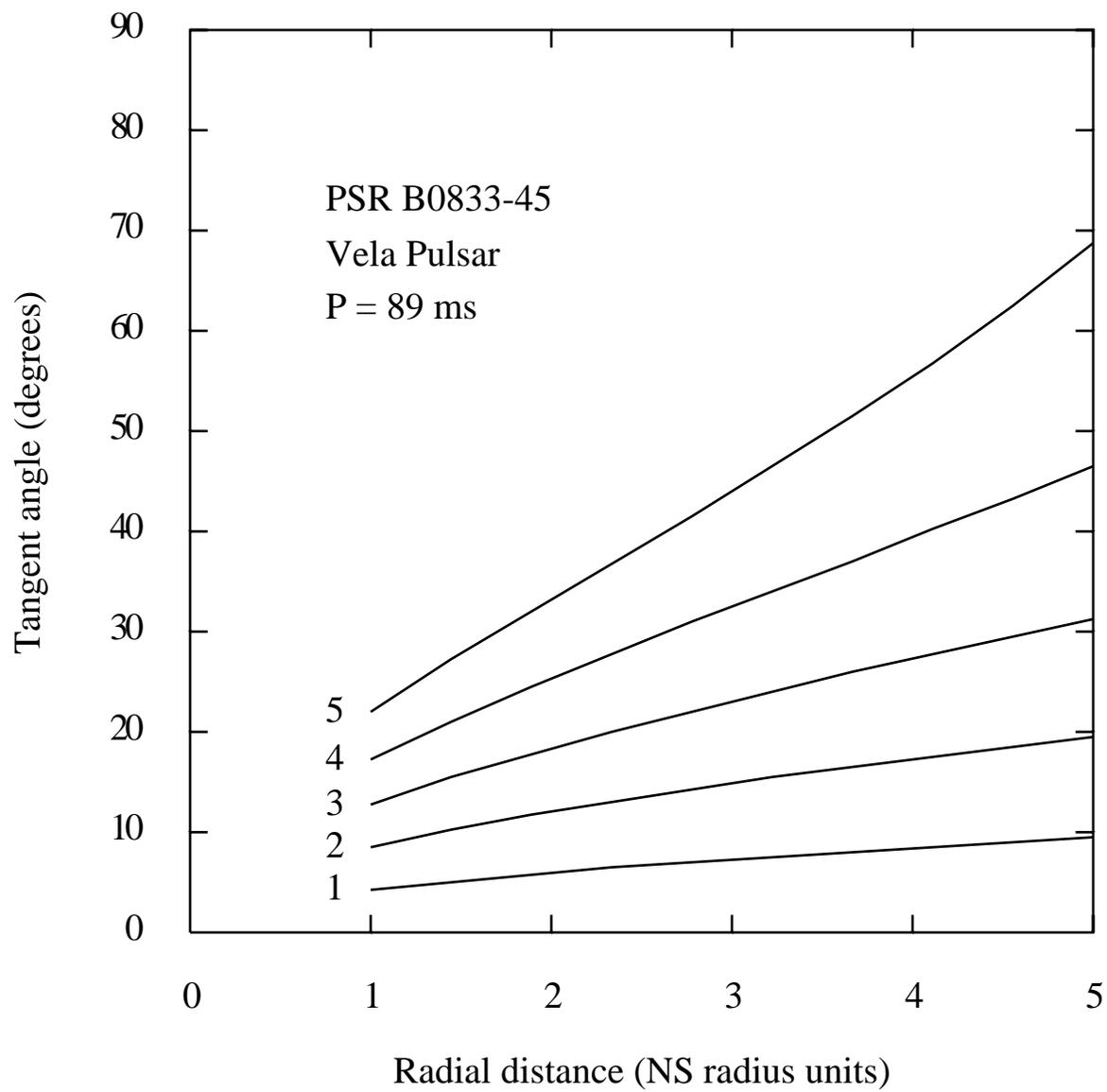
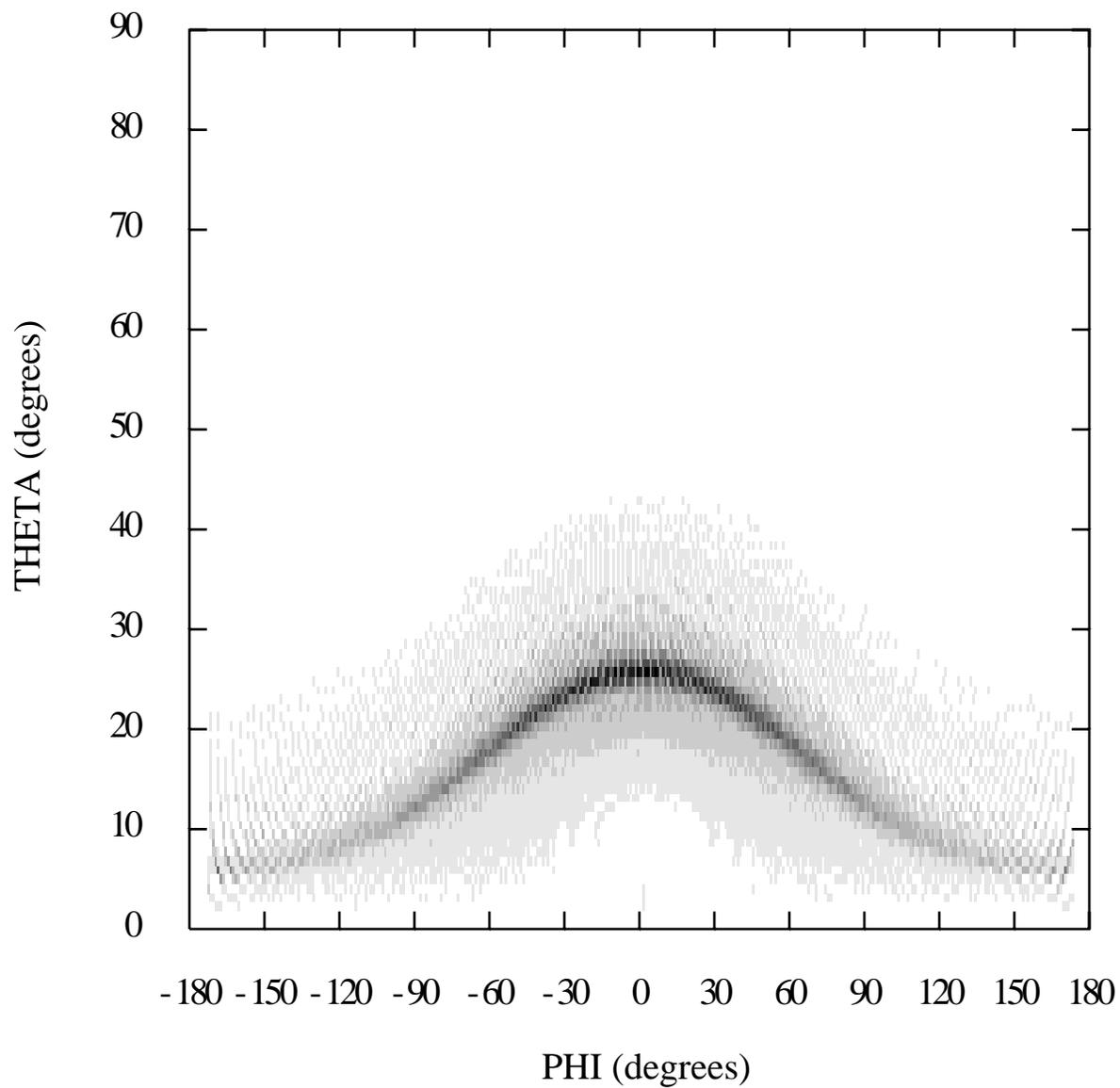
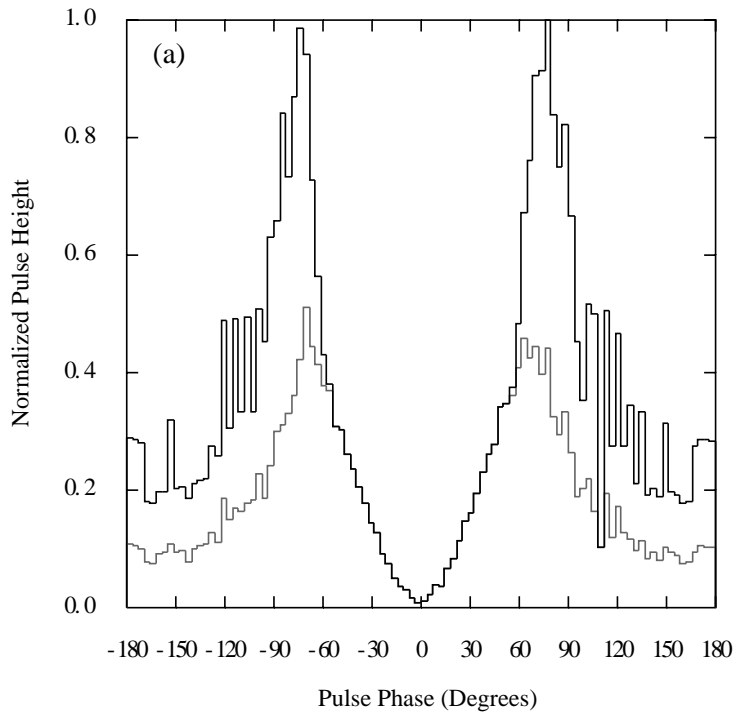
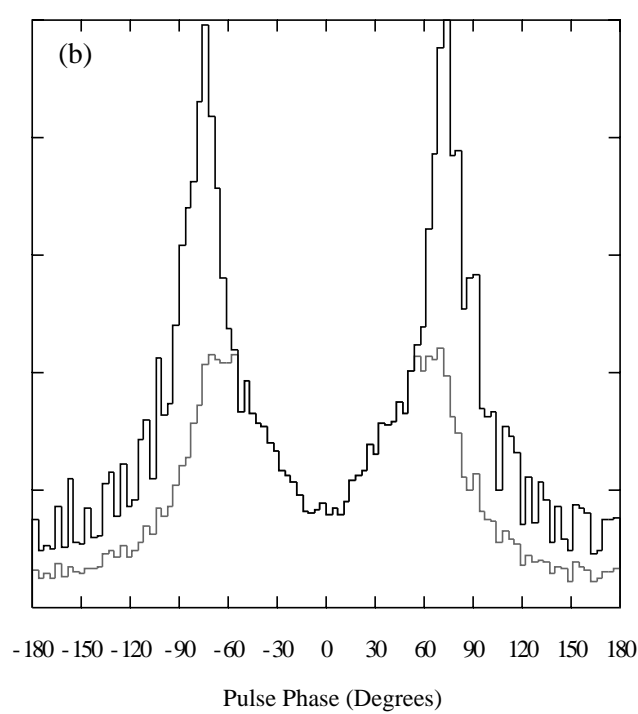
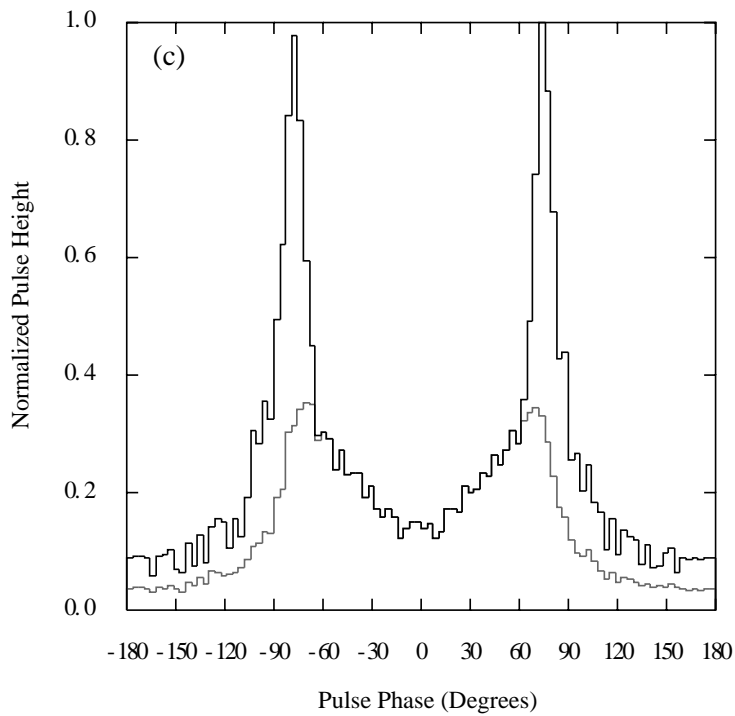
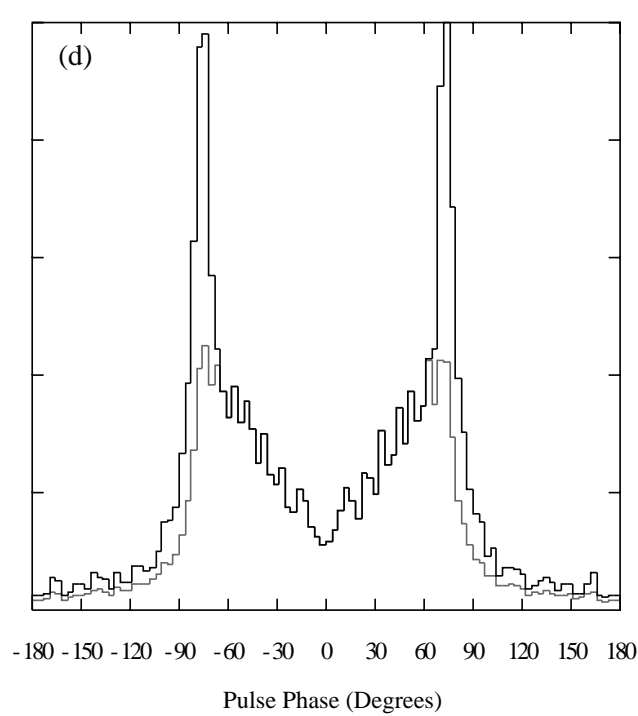
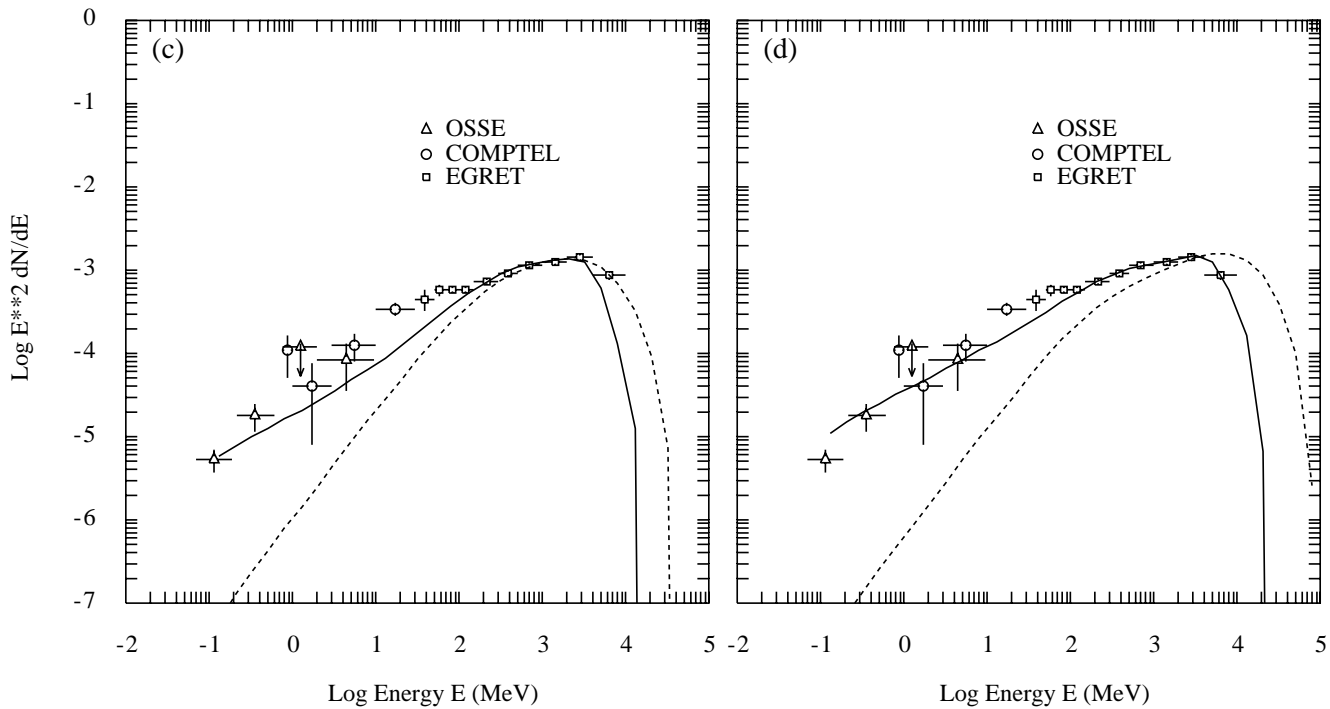
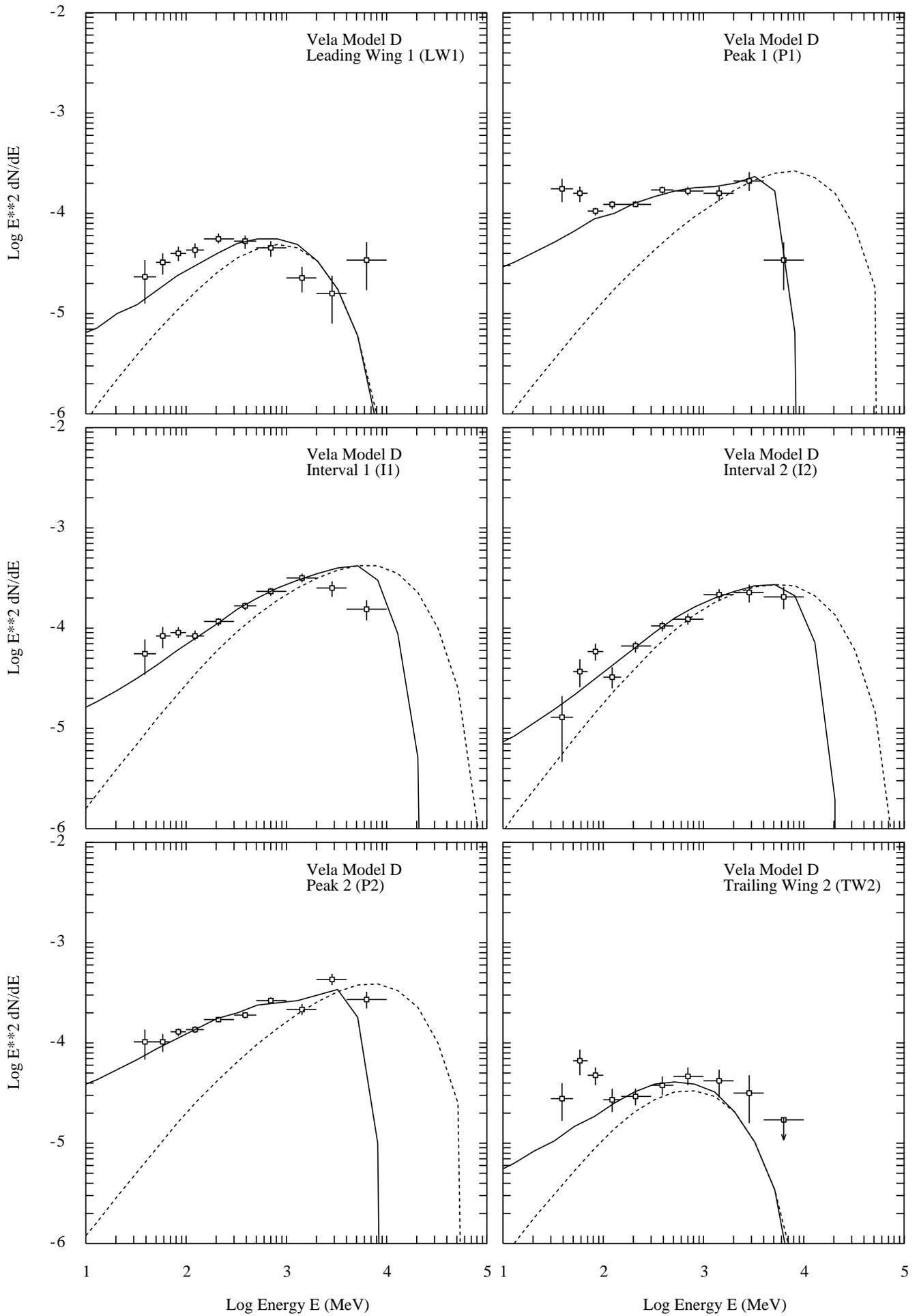
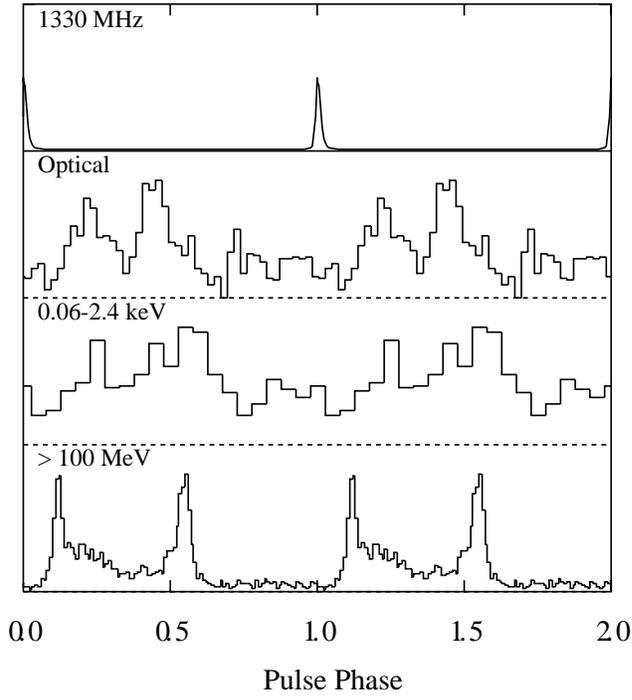
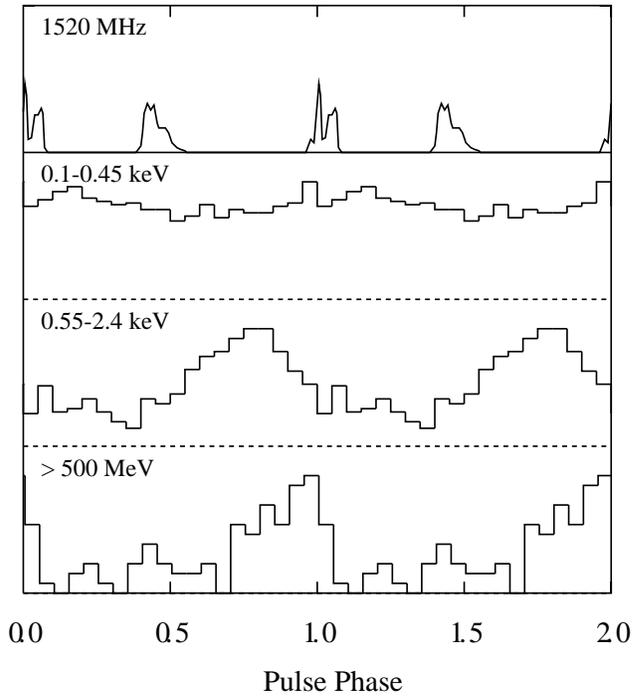
\end{document}